# The influence of additives in the stoichiometry of hybrid lead halide perovskites

Ignasi Burgués-Ceballos,[1] Achilleas Savva,[1] Efthymios Georgiou,[1] Konstantinos Kapnisis,[1] Paris Papagiorgis,[2] Androniki Mousikou,[2] Grigorios Itskos,[2] Andreas Othonos,[3] and Stelios A. Choulis[1,a]

[1]*Molecular Electronics and Photonics Research Unit, Department of Mechanical Engineering and Materials Science and Engineering, Cyprus University of Technology, 45 Kitiou Kyprianou Street, Limassol 3041, Cyprus*

[2]*Experimental Condensed Matter Physics Laboratory, Department of Physics, University of Cyprus, Nicosia 1678, Cyprus*

[3]*Laboratory of Ultrafast Science, Department of Physics, University of Cyprus, Nicosia 1678, Cyprus*

a)  Corresponding author: stelios.choulis@cut.ac.cy

We investigate the employment of carefully selected solvent additives in the processing of a commercial perovskite precursor ink and analyze their impact on the performance of organometal trihalide perovskite ($CH_3NH_3PbI_{3-x}Cl_x$) photovoltaic devices. We provide evidence that the use of benzaldehyde can be used as an effective method to preserve the stoichiometry of the perovskite precursors in solution. Benzaldehyde based additive engineering shows to improve perovskite solid state film morphology and device performance of $CH_3NH_3PbI_{3-x}Cl_x$ based solar cells.

**THE MANUSCRIPT**

Earth-abundant organic-inorganic trihalide perovskites possess excellent optoelectronic properties and have been successfully implemented in a wide range of applications, including light emitting diodes, sensors and photovoltaic devices.[1] Their general structure is $ABX_3$, where A is an organic cation, B a divalent metal ion and X a Cl, Br, I or any mixture of these halides. The countless composition combinations offer the possibility of tuning important properties such as the energy band gap or the stability of the resulting photoactive layers.[2] Indeed, compositional engineering is one of the main strategies used to boost device performance.[3–5] High quality perovskite layers can be obtained by properly controlling the crystal growth.[6,7] However, some of the most commonly used precursors as well as the resulting perovskite layers tend to decompose upon exposure to humidity and high temperature, even under material encapsulation.[8] Unstable perovskite precursors impede a precise control of the stoichiometry to yield high quality films. This is even more severe in the case of mixed halides, where the ratio of the precursors can be critical.[9] On the other hand, solvent engineering and the use of additives have also contributed meaningfully to the field.[10–14]

In this work we examine the effect of solvent additives engineering on the morphology, optoelectronic properties and solar cell device performance of solution processed methylammonium lead mixed iodide-chloride perovskites ($CH_3NH_3PbI_{3-x}Cl_x$) photovoltaics. Noteworthy, we use a commercial precursor ink containing methylammounium iodide (MAI), $PbCl_2$ and $PbI_2$ at a molar ratio of 1:1:4 ($PbCl_2$:$PbI_2$:MAI) in anhydrous N,N-dimethylformamide (DMF). Importantly, in order to produce a representative study we employ a set of the following criteria to select the appropriate solvent additives: i) inclusion of both higher and lower polarity solvents compared to the reference host DMF solvent, ii) use of higher boiling point (bp) solvents than DMF (bp 153 °C), while iii) keeping high enough vapour pressure to ensure a complete solvent removal during perovskite film formation at low temperature (80 °C) and iv) different nucleophilic strength and/or reducing character. The selected additives are: 1,2,3,4-tetrahydronaphthalene (*aka* tetralin), γ-valerolactone, acetophenone, furfural, benzaldehyde, N-methylformamide, dimethyl sulfoxide (DMSO) and benzylamine (see Table I).

TABLE I. Solvent additives used in this study.

| Solvent | Polarity[a] $\delta P$ (MPa$^{1/2}$) | Boiling point (°C) | Vapor Pressure, 25 °C (mm Hg) | Reducing character |
|---|---|---|---|---|
| DMF[b] | 13.7 | 153 | 3.68 | Weak |
| Tetralin | 2.0 | 207 | 0.35 | Very weak |
| γ-Valerolactone | 11.5 | 186 | 0.55 | Weak |
| Acetophenone | 9.0 | 202 | 0.30 | Weak |
| Furfural | 14.9 | 162 | 1.00 | Weak |
| Benzaldehyde | 7.4 | 179 | 1.19 | Medium |
| N-methylformamide | 18.8 | 200 | 0.16 | Weak |
| DMSO | 16.4 | 189 | 0.49 | Weak |
| Benzylamine | 4.6 | 185 | 0.57 | Strong |

[a] The energy from dipolar intermolecular force between molecules, from Hansen Solubility Parameters.
[b] DMF is the host solvent

A detailed description of the perovskite film preparation, device fabrication[15] and the morphological and optical characterization can be found in the supplementary material. As shown in Fig. 1, the addition of small amounts (1-5% vol) of these solvents into the precursor ink yielded a set of perovskite layers with distinct morphological characteristics. Some of the additives used, namely tetralin, acetophenone, γ-valerolactone and N-methylformamide, do not seem to have a clear influence on crystal size. On the other hand, scanning electron microscopy (SEM) revealed significant modifications in crystal size, shape and surface coverage (Fig. 1) upon the inclusion of the other additives. The case of DMSO deserves a special mention, as its ability to form the PbI$_2$(DMSO)$_2$ complex has been used to retard the crystallization process and modify the crystal size.[10,16,17] A dense layer of small perovskite crystals was obtained with 1% vol. of DMSO, larger crystals and formation of pinholes were visible at the level of 3% vol. additive while incorporation of 5% vol of DMSO yielded significantly larger crystals with a very poor surface coverage. The wide variation in the microstructure within the small range of DMSO additive employed provides evidence for the critical role of the amount of additives in the structural properties of the perovskite solids. Similarly, the crystal size increased with increasing additive content for benzaldehyde and furfural, though in this case the dense film packing appears unaffected by the additive treatment. Importantly, the addition of only 1% vol. of benzylamine resulted in the formation of large cuboid-shaped crystals, which could be related to the changes in stoichiometry discussed later in the text. UV-Vis absorbance and relevant Tauc plots indicate that the additive treatment has a minor (maximum shift of 5 meV compared to the reference DMF solvent) and non-systematic influence on the material energy band gap, with the exception of benzylamine where a ~20 meV red-shift is observed (Fig. S1). This moderate change could be related to lattice structural distortions derived from a modified stoichiometry in the perovskite film formation.[18] A more accurate X/Pb stoichiometric ratio would reduce the amount of defects in the perovskite layer.

The photovoltaic efficiency measurements under calibrated AM 1.5G irradiation conditions performed for several identical experimental device fabrication runs reveal a consistent trend (Fig. 2): the maximum power conversion efficiency (PCE) corresponds systematically to the devices prepared with the benzaldehyde additive. Regardless of its content, the use of benzaldehyde was beneficial for the device performance, systematically improving by a factor of 10% the PCE of the control devices (from an average of 11% to 12%), with a 13.5% record device. The improvement appears to be mostly a consequence of an increased photocurrent, with a minor enhancement in the fill factor (Fig. S2). Furthermore, the initial stability test of those solar cells is increased by more than a threefold factor i.e. from 200 h to 650 h of operational lifetime ($T_{80}$) under the ISOS-D-1 protocol. We attribute the improvements to a better perovskite solid state film formation induced by the benzaldehyde treatment. The microscopy data shown within Fig. 1 provides indication that the treatment with benzaldehyde improves the perovskite active layer morphology compared to the untreated original ink.

In the $CH_3NH_3PbI_{3-x}Cl_x$ based formulation under investigation we did not observe any clear correlation between the perovskite film morphology nor the device performance with the polarity, boiling point and vapor pressure of the additives tested. Even so, the most significant changes described above were obtained with DMSO, benzylamine and benzaldhyde, three solvents whose boiling points are 25-35 °C above that of the host solvent DMF. Therefore we can not exclude the relevance of this parameter. Moreover, we believe that further studies based on the selection of additives presented here are necessary to fully understand their morphological influence of polarity, boiling point and vapor pressure on alternative perovskite formulations.

Based on the results of our preliminary study, the parameter which appears to have a measurable impact on the film morphology and device performance is the reducing character of the additive. Thus we further explored how the additive reducing character affects the film characteristics and in particular the stoichiometry of the perovskite material. For that reason we characterized the absorbance of the precursor formulations with increasing additive content (Fig. 3). Firstly, the focus was set on the spectral features at 320 nm and 365 nm, corresponding to absorption signatures of $PbI_2$ and MAI, respectively.[19] It is well known that organic iodides such as MAI are not very stable in air and under light irradiation, mainly due to the oxidation of $I^-$ into $I_2$. By monitoring the $PbI_2$ and MAI features, we aimed to probe potential variations in the stoichiometry of the original precursor ink induced by the additives. The resulting spectra showed no changes when using tetralin, γ-valerolactone, DMSO or N-methylformamide additives. On the other hand, the $PbI_2$ signal was slightly stronger (weaker) with acetophenone and furfural (benzylamine). Finally, both $PbI_2$ and MAI peaks showed a meaningful increase upon benzaldehyde treatment. We ascribe this effect to the reducing character of benzaldehyde, which appears to prevent the oxidation of the unstable precursors and eventually reduce already oxidized components. According to table I, benzylamine is a stronger reducing agent and thus one could expect it to undergo nucleophilic subsitution with the methylammonium cation, yielding benzylammonium and thus lowering even further the concentration of MAI. However, as shown in Fig. 3 the MAI peak does not diminish upon addition of benzylamine, whereas the PbI$_2$ signal at 320 nm decreases instead. We could not confirm whether this is due to the complexation of Pb with the amine groups. The experimental results however suggest that with a low amount (1% vol) of benzylamine only the PbI$_2$ undergoes undesired reactions, while the PbCl$_2$ remains unaffected and favor the growth of cuboid shaped perovskite crystals[6] (see bottom right corner in Fig. 1) instead of the tetragonal-cubic mixture expected in mixed iodide-chloride perovskites. This is because of the lower ionization energy of iodine (10.45 eV) compared with chlorine (12.97 eV). At higher concentrations, the excess amine content might also affect the PbCl$_2$, hence the formation of MAI crystallites (top right corner in Fig. 1). Such modification in the perovskite crystal structure is plausible origin of the bathochromic shift of the absorption edge observed upon treatment with benzylamine (Fig. S1), as discussed previously in the manuscript.

On the other hand, the milder influence of benzaldehyde appears not to alter the crystal structure and at the same time reveals more clear trends in the signatures associated with the different oxidation states of iodine (Fig. 3b): the amount of iodide ion I$^-$ increases (shoulder at 193 nm and peak at 248 nm) in relation to I$_2$ (203 nm), whereas the content of the partially oxidized triiodide ion I$_3^-$ decreases (288 nm and 350 nm). It is therefore proven that the beneficial effect of adding benzaldehye in the precursor solution stands for its reducing properties. Importantly, the fact that this change is observed in solution supports the idea that this is an isolated effect rather than a combined action with the boiling point or the vapor pressure, which one could expect to have an impact on the drying kinetics during film formation. This valuable finding demonstrates that the limiting instability of the perovskite precursors can be significantly mitigated by the simple addition of small amounts of specific reducing agents such as the herein proposed benzaldehyde. As a result, the modified ink has a higher unreacted precursor content compared to the untreated material, which results in the improved perovskite microstructure.[20] Our results are in good agreement with the work reported by Zhang et al., where the addition of hypophosphorous acid as a reducing agent yielded a significantly enhanced perovskite microstructure.[14] In our case, we believe that the relatively low precursor concentration of the untreated commercial ink is responsible for the lower solar cell efficiencies achieved with such a material compared to the reported state of the art devices. Furthermore, in contrast to the aforementioned work[14] we do not observe a consistent enhancement of the film PL intensity nor PL lifetime with perovskite crystal size upon addition of the reducing additive, as observed in Fig. S3. This may be understood as the measured PL lifetime is a convolution of non-radiative quenching channels and the radiative decay term that are potentially affected by the additive treatment in a competing and complex way. Similarly, Liang et al reported a PL lifetime quenching in perovskite films treated by additives, as a result of the enhanced crystallization and the more efficient charge transfer to the PEDOT interlayer.[13] We additionally observe that the MAI to $PbI_2$ ratio increases upon benzaldehyde addition. The role of non-stoichiometric precursor ratios is still a controversial topic.[21,22] In our work, the shifted stoichiometry, with a slight excess of MAI compared to the original ink, might also be a contributing factor for the formation of improved quality perovskite films. The addition of this solvent appears therefore as a simple way to control the $CH_3NH_3PbI_{3-x}Cl_x$ perovskite morphology and stoichiometry and yield better solar cell device performances for the $CH_3NH_3PbI_{3-x}Cl_x$ precursor ink.

In summary, we have investigated the use of solvent additives in a commercial $CH_3NH_3PbI_{3-x}Cl_x$ ink to modify the morphology and optoelectronic properties of the resulting perovskite layers. We have observed a diverse impact on such properties depending on the additive used. Crystal size and surface coverage can suffer large variations upon additive treatment. Although our results suggest no direct influence of the solvent polarity, boiling point and vapor pressure on the properties and device performance of the $CH_3NH_3PbI_{3-x}Cl_x$ material used for this investigation, we beleive that the selection process of the additives used within this paper can be interested for further studies on alternative perovskite formulations. In contrast, we provide evidence that some of these additives can effectively modify the precursor concentration and stoichiometry in the original perovskite ink. The experimental results indicate that the addition of a low amount (1% vol) of benzylamine favor the growth of $CH_3NH_3PbI_{3-x}Cl_x$ cuboid shaped perovskite crystals instead of the tetragonal-cubic mixture expected in mixed iodide-chloride perovskites, due to the lower ionization energy of iodine versus chloride and the consequent faster interaction of the additive with $PbI_2$. For the tetragonal-cubic mixture $CH_3NH_3PbI_{3-x}Cl_x$ perovskites studied benzaldehyde showed the strongest beneficial morphological effect, which we ascribe mainly to its mild reducing character. As a result the power conversion efficiency of $CH_3NH_3PbI_{3-x}Cl_x$ solar cell devices using addition of small amount (1-5% vol) of benzaldehyde additive increased by 10% and the accelerated lifetime performance under ISOS-D-1 conditions is improved.

**Supplementary Material**

Supplementary information includes details for materials, processing of perovskite films and fabrication of perovskite devices. Additional information for morphological and optical characterization of perovskite films and precursor formulations as well as perovskite solar device performance at different additives concentrations to show the consistency of the reported results is included.

**Acknowledgements**

This project has received funding from the European Research Council (ERC) under the European Union's Horizon 2020 research and innovation programme (grant agreement No 647311).

**FIGURE AND FIGURES CAPTIONS**

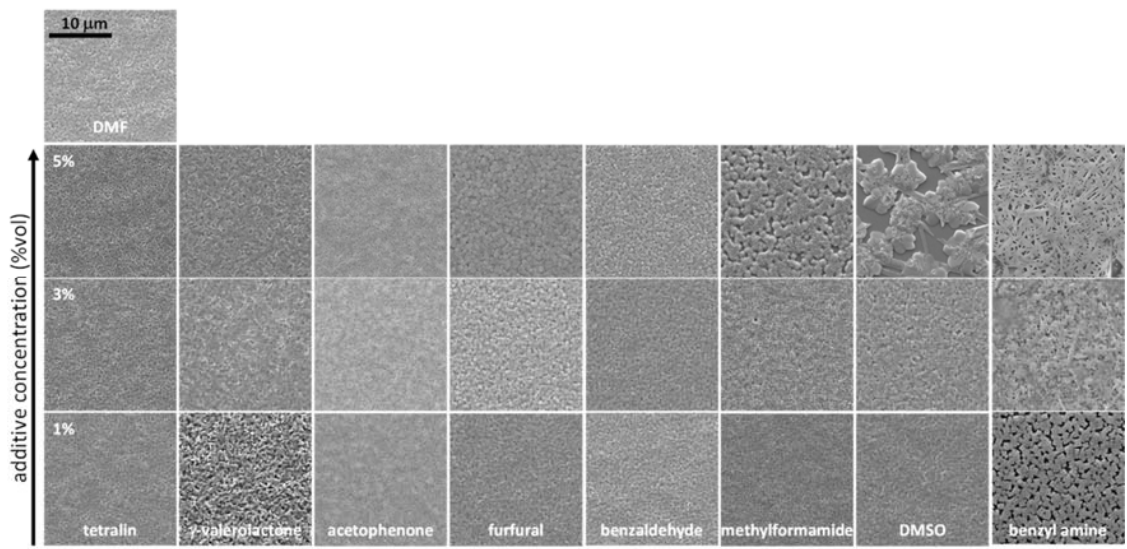

FIG. 1. Morphology of $CH_3NH_3PbI_{3-x}Cl_x$. SEM images of films prepared from precursor solutions containing different additives.

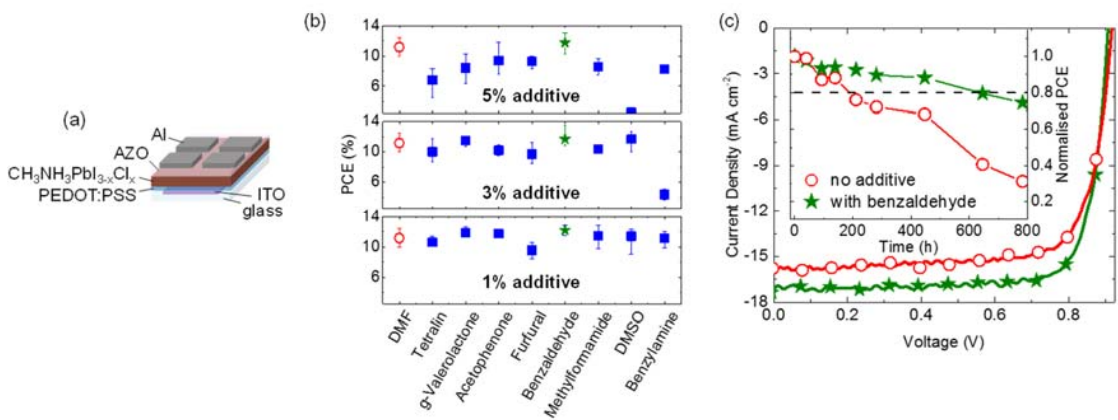

FIG. 2. Impact of the additives on the photovoltaic performance and stability. (a) Schematic of layer stacking of the perovskite-based solar cell. (b) Power conversion efficiency as a function of the additive and its concentration. (c) Representative current density vs voltage characteristics of devices without additive (control) and with 5% vol. of benzaldehyde. Inset shows lifetime data under ISOS-D-1 protocol.

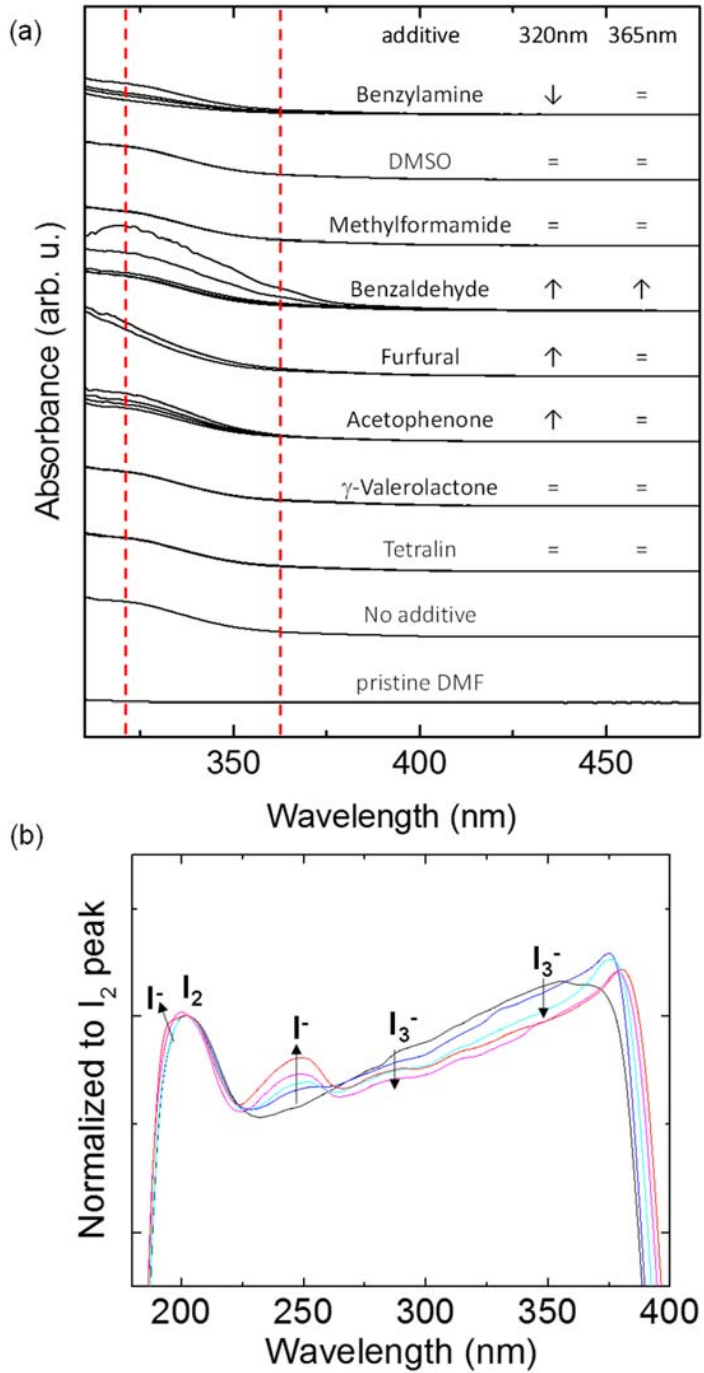

FIG. 3. Absorbance spectra of perovskite precursor solutions. The measurements were performed on formulations containing different additive content. The caption summarizes the changes observed in (a) the signals at 320 nm and 365 nm (marked with vertical dashed lines), which correspond to signatures of $PbI_2$ and MAI, respectively[14,19] and (b) the changes assigned to different oxidation states of iodine upon addition of benzaldehyde.

# Supplementary Material

**The Influence of Additives in the Stoichiometry of Hybrid Lead Halide Perovskites**

Ignasi Burgués-Ceballos,[1] Achilleas Savva,[1] Efthymios Georgiou,[1] Konstantinos Kapnisis,[1] Paris Papagiorgis,[2] Androniki Mousikou,[2] Grigorios Itskos,[2] Andreas Othonos,[3] and Stelios A. Choulis[1,a]

[1]*Molecular Electronics and Photonics Research Unit, Department of Mechanical Engineering and Materials Science and Engineering, Cyprus University of Technology, 45 Kitiou Kyprianou Street, Limassol 3041, Cyprus*

[2]*Experimental Condensed Matter Physics Laboratory, Department of Physics, University of Cyprus, Nicosia 1678, Cyprus*

[3]*Laboratory of Ultrafast Science, Department of Physics, University of Cyprus, Nicosia 1678, Cyprus*

a) Corresponding author: stelios.choulis@cut.ac.cy

**Materials**

All the perovskite-based films and devices were prepared using a commercial precursor ink (I201) from Ossila Ltd. The formulation contains methylammounium iodide (MAI), $PbCl_2$ and $PbI_2$ at a molar ratio of 1:1:4 ($PbCl_2$:$PbI_2$:MAI) in anhydrous DMF solvent. The rest of solvents used were purchased from Sigma-Aldrich and used as received. Pre-patterned glass-ITO substrates (sheet resistance 4 $\Omega$/sq) were purchased from Psiotec Ltd, PEDOT:PSS from H.C. Stark (Clevios P VP Al 4083), $PC_{70}BM$ from Solenne BV and aluminum-doped zinc oxide (AZO) ink (N-20X) from Nanograde.

**Fabrication of $CH_3NH_3PbI_{3-x}Cl_x$ perovskite films**

ITO substrates were cleaned by successive steps using an ultrasonic bath in acetone and isopropanol for 10 min each and then heated at 120 °C on a hot plate for 10 minutes immediately prior to use. A 40 nm PEDOT:PSS layer was formed by dynamically spin coating the ink at 6000 rpm for 30s in air. The substrates were then transferred to a nitrogen filled glove box for a 20 min annealing at 120 °C. Small quantities of additives (1%, 3% or 5% vol.) were added to the commercial perovskite precursor formulation and stirred for 5 min prior to use. The perovskite layer was formed on top of PEDOT:PSS by dynamically spin coating the modified ink at 4000 rpm for 30 s and subsequent annealing for 2h at 80 °C, resulting in a ca. 350 nm thick films.

**Fabrication of $CH_3NH_3PbI_{3-x}Cl_x$ perovskite solar cells**

A more detailed description of the solar cell architecture used in this work can be found elsewhere.[15] The whole device structure was ITO/PEDOT:PSS/$CH_3NH_3PbI_{3-x}Cl_x$/$PC_{70}BM$/AZO/Al. To complete the solar cell stack, a 50 mg/mL solution of $PC_{70}BM$ in chlorobenzene was dynamically spin coated on the perovskite layer at 1000 rpm for 30s. On top of it, the AZO ink was dynamically spin coated at 1000 rpm for 30s to yield ca. 50 nm thick films, which were then annealed at 80 °C for 2 min. Finally, 100 nm of Al were thermally evaporated through a shadow mask to define the active area (9 mm$^2$) of the devices. A UV-activated epoxy resin (E131 from Ossila) was used to encapsulate the devices with glass coverslips.

**Morphological and optical characterization of perovskite films and precursor formulations**

The thicknesses of the layers were determined with a Veeco Dektak 150 profilometer. Scanning Electron Microscopy (SEM) measurements were carried out with a Quanta 200 (FEI, Hillsboro, Oregon, USA) at various magnifications, 10 kV beam voltage and 3.0 spot size. The SEM images were used to correlate morphological features with the additive characteristics. Absorbance of the precursor formulations with different additive content were characterized with a Shimadzu UV-2700 UV-Vis spectrophotometer. Steady-state PL experiments were performed in a Fluorolog- 3 Horiba Jobin Yvon spectrometer based on an iHR320 monochromator equipped with a visible photomultiplier tube (Horiba TBX-04 module). The PL was non-resonantly excited at 375 nm by the line of 5 mW Oxxius laser diode. Time-resolved PL was measured using a time correlated single photon counting (TCSPC) method at the same spectrophotometer. The PL was excited by a NanoLED laser diode at 375 nm operating at 100 KHz with a pulse FWHM of ~100 ps. The PL decays were analyzed and line-fitted using the DAS6 (Horiba Jobin Yvon) software analysis package. All PL data were acquired with samples placed in vacuum (~$10^{-3}$ mbar) in a custom-made optical vacuum chamber.

**Perovskite solar cell characterization**

Current versus voltage characteristics were measured with a LIV Functionality Test System (Botest Systems GmbH) under calibrated AM 1.5G irradiation on a Newport solar simulator (100 mW/cm$^2$). Both forward (short circuit → open circuit) and reverse (open circuit → short circuit) scans were measured with 10 mV voltage steps and 40 ms of delay time. A custom made shadow mask was used during the measurements to accurately define the device area. The ISOS-D-1 protocol was followed to carry out the stability characterization.

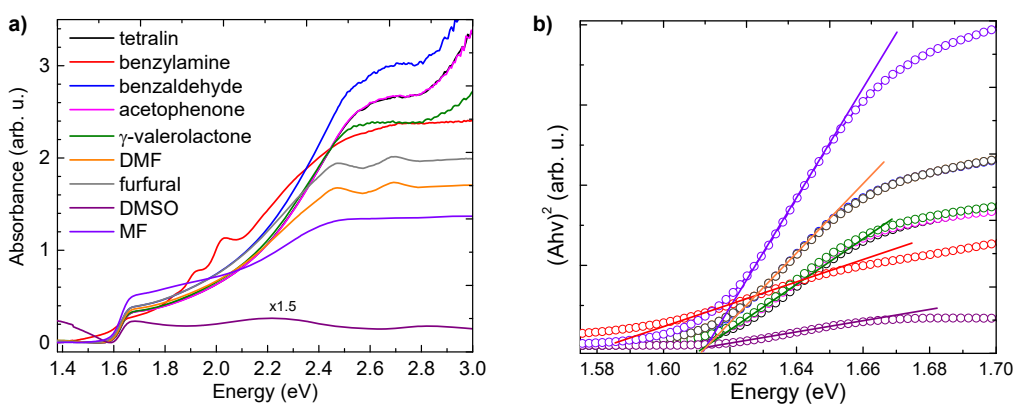

*Figure S1. a) UV-Vis absorbance spectra and b) Tauc plots of perovskite films prepared from precursor solutions containing different additives.*

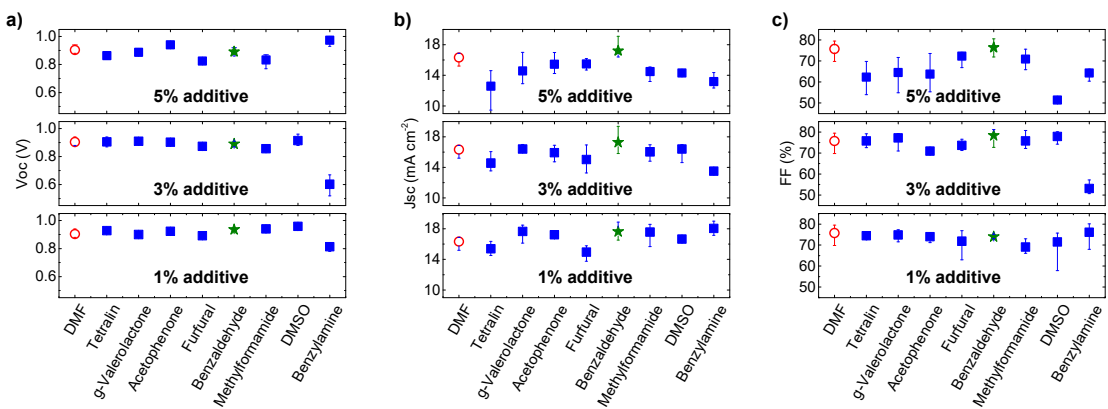

*Figure S2. Influence of additive and concentration on a) open circuit voltage, b) short circuit current density and c) fill factor. The control devices (100% DMF) are marked with open circles. The best performing devices (with benzaldehyde additive) are pointed with stars.*

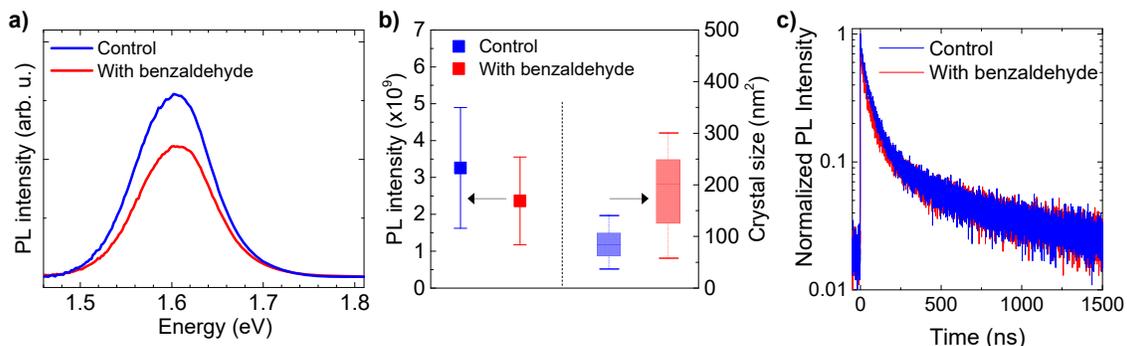

*Figure S3. a) Photoluminescence (PL) spectra, b) average PL intensity vs. crystal size and c) PL transient decays of perovskite films deposited on top of PEDOT:PSS from the original ink (control) and with the addition of 5% vol. of benzaldehyde. Triple exponential fits on the transient decays gave an average PL lifetime of 103 ns and 82 ns, respectively.*